\documentclass[twocolumn,prl]{revtex4}
\usepackage{graphicx}
\usepackage{dcolumn}
\usepackage{epsfig}
\begin{document}
\title{Localized activity profiles and storage capacity of rate-based autoassociative networks}
\author{Yasser Roudi$^{\S}$ and Alessandro Treves$^{\S,\P}$\\
\small $^{\S}$ Scuola Internazionale Superiore di Studi Avanzati,\\ Settore di Neuroscienze Cognitive, Trieste, Italy \\
\small $^{\P}$ NTNU, Centre for the Biology of Memory, Trondheim,
Norway}

\begin{abstract}

We study analytically the effect of metrically structured
connectivity on the behavior of autoassociative networks. We focus
on three simple rate-based model neurons: threshold-linear, binary
or smoothly saturating units. For a connectivity which is short
range enough the threshold-linear network shows localized
retrieval states. The saturating and binary models also exhibit
spatially modulated retrieval states if the highest activity level
that they can achieve is above the maximum activity of the units
in the stored patterns. In the zero quenched noise limit, we
derive an analytical formula for the critical value of the
connectivity width below which one observes spatially non-uniform
retrieval states. Localization reduces storage capacity, but only
by a factor of $2\sim 3$. The approach that we present here is
generic in the sense that there are no specific assumptions on the
single unit input-output function nor on the exact connectivity
structure.
\end{abstract}

\maketitle

Recurrent neuronal networks, when endowed with associative
synaptic plasticity, are able to learn patterns of activity and
retrieve them later when provided with partial cues -- a property
called autoassociative retrieval. This is believed to be an
important ability of neocortical, as well as of hippocampal
networks \cite{Rol98}. In the hippocampus, where several
experiments are currently aimed at demonstrating attractor
dynamics \cite{Wil+05}, it is the CA3 subfield which is thought to
operate as a recurrent autoassociative memory, and its recurrent
connections could be modeled, to a crude approximation, as
extending uniformly across the network. In the neocortex, instead,
the metrical organization of the connectivity cannot be neglected:
neurons close to each other in the cortex are much more likely to
be connected, while this probability becomes very low with
distance \cite{Bra91}. Even in simplified theoretical models,
there are technical problems that complicate the analysis of
associative networks with non-uniform connectivity: the distance
dependence in the connectivity forces one to introduce "field"
order parameters in the model\cite{Rou04}; moreover, asymmetric
connectivity makes unapplicable those methods of equilibrium
statistical mechanics which were originally used for solving the
classical models of associative retrieval \cite{Ami89}.

Recently there have been several studies on how structure in the
connectivity affects performance of recurrent
networks\cite{Tor04,Mor04,Ana05,Rou04,Kor05}. Most of these
studies approach the problem by means of simulations and by
focusing on one particular model neuron and connectivity
structure. In one of these studies \cite{Rou04} we described an
analytical treatment of the problem in the case of an associative
network with threshold-linear ({\em TL}) units. Due to
difficulties in solving the steady-state equations, we developed
an approximate method for calculating the storage capacity of such
a network, which was in reasonable agreement with simulations. In
this paper, instead, we derive the equations that govern the
steady-state properties of a generic rate-based associative
network and introduce a numerical method that can be used for
solving these equations with arbitrary accuracy. We calculate
accurately the results that we earlier obtained approximately for
the {\em TL} network and extend the same analysis to a network of
binary units as well as to a model which includes firing rate
saturation. As a result of short range connectivity, in the {\em
TL} network retrieval states may appear as localized bumps of
activity, while these localized solutions are absent in a network
with $0-1$ binary units. The possibility of spatial modulation for
retrieval states depends on the maximum rate of the units relative
to their rates in the memory patterns, and on how this saturation
rate is approached.

Consider a network of $N=2L \rightarrow \infty$ units, in which
the firing rate of the unit located at position $\textbf{r}$ is
represented by a variable $v(\textbf{r})\geq 0$. We assume that
each unit receives $C \rightarrow \infty$ inputs from the other
units in the network. The 'Hebbian' learning rule we consider
prescribes that the synaptic weight between units $\textbf{r}$ and
$\textbf{r}'$ be given as:
\begin{equation}
J(\textbf{r},\textbf{r}')=\frac{1}{Ca^2}\sum_{\mu=1}^p
\wp(\textbf{r};\textbf{r}')\left(\eta^{\mu}(\textbf{r})-a\right)
\left(\eta^{\mu}(\textbf{r}')-a\right),
\end{equation}
where $\eta^{\mu}(\textbf{r})$ represents the activity of unit
located at $\textbf{r}$ in memory pattern ${\mu}$ and $\wp
(\textbf{r};\textbf{r}')$ is a function depending just on
$|r_i-r'_i| \forall i$ which gives the probability that two
neurons one at $\textbf{r}$ and the other at $\textbf{r}'$ are
connected to each other ($r_i$ is a component of \textbf{r}).

Each $\eta^{\mu}(\textbf{r})$ is taken to be a `quenched variable'
drawn independently from a distribution $p(\eta)$, with the
constraints $\eta \ge 0$, $\langle \eta \rangle = \langle \eta^2
\rangle= a$, where $\langle\rangle$ stands for the average over
the distribution $p(\eta)$ \cite{Rol98}. Here we concentrate on
the binary coding scheme
$p(\eta)=a\delta(\eta-1)+(1-a)\delta(\eta)$, but the calculation
can be easily extended to any probability distribution. We further
assume that the input (local field) to the unit located at
\textbf{r} is given by:
\begin{equation}
h(\textbf{r})=\int d\textbf{r}' J(\textbf{r};\textbf{r}')
v(\textbf{r}')+ b\left(\{x_k\}\right), \label{locfield}
\end{equation}
where the first term enables the memories encoded in the weights
to determine the dynamics. In the second term,
$x_k=X_k[v(\textbf{r})]$ are variables which depend on the global
properties of the network activity. An example is the mean
activity of the network $x\equiv \frac{1}{N}\int d\textbf{r}
v(\textbf{r})$ which, through the second term, might be used to
regulate itself, so that at any moment in time it approaches {\em
sparsity} $a$. Following \cite{Rou04}, we start our analysis by
defining as an order parameter the \emph{local overlap}:
\begin{equation}
m^{\mu}(\textbf{r})=\frac{1}{C}\int d\textbf{r}'
\wp(\textbf{r}';\textbf{r})
(\eta^{\mu}(\textbf{r}')/a-1)v(\textbf{r}').
\end{equation}
The pattern $\nu$ is said to be retrieved if $\int d\textbf{r}
m^{\nu}(\textbf{r})=O(N)$. We denote as $m(\textbf{r})\equiv
m^{\nu}(\textbf{r})$ the local overlap with the pattern $\nu$ to
be retrieved by a partial cue.

The activity of each unit is determined by its input-output
transfer function $v(\textbf{r})=F(h(\textbf{r}))$. In general,
using a self-consistent signal-to-noise analysis \cite{Shi92,
Rou04}, the fixed-point equations for such a network reduce to:
\begin{eqnarray}
\psi(\textbf{r}_2;\textbf{r}_1)&=&\int d\textbf{r} K(\textbf{r}_2;\textbf{r}) \wp(\textbf{r}';\textbf{r}_1)\nonumber\\
&&\space+\int d\textbf{r}d\textbf{r}' K(\textbf{r}_2;\textbf{r}) K(\textbf{r};\textbf{r}')\wp(\textbf{r};\textbf{r}_1)+\dots \nonumber \\
K(\textbf{r}_2;\textbf{r})&=&\frac{T_0}{C}\wp(\textbf{r}_2;\textbf{r})\langle\int Dz G'[\textbf{r}]\rangle \equiv \wp (\textbf{r}_2;\textbf{r})\Phi(\textbf{r})\nonumber \\
\Gamma(\textbf{r})&=&\alpha T_0 \psi(\textbf{r};\textbf{r})\nonumber\\
\rho^2(\textbf{r}_2)&=&\frac{\alpha T_0^2}{C}\int d\textbf{r}_1 A(\textbf{r}_2;\textbf{r}_1) I_3(\textbf{r}_1) \nonumber \\
m(\textbf{r}_2)&=&\frac{1}{C}\int d\textbf{r}_1 \wp(\textbf{r}_2;\textbf{r}_1) I_2(\textbf{r}_1)\nonumber \\
x_k&=&X_k[\int Dz G[\textbf{r};\Gamma]]
\end{eqnarray}
where $\alpha=p/C$ is the storage load and:
\begin{eqnarray}
I_2(\textbf{r})&=&\langle(\eta({\textbf{r}})/a-1)\int Dz G[\textbf{r};\Gamma]\rangle\nonumber\\
I_3(\textbf{r})&=&\langle\int Dz G[\textbf{r};\Gamma]^2\rangle \nonumber\\
A(\textbf{r}_2;\textbf{r}_1)&=&\wp(\textbf{r}_2;\textbf{r}_1)+2\wp(\textbf{r}_2;\textbf{r}_1)\psi(\textbf{r}_2;\textbf{r}_1)\nonumber\\
&&+\psi(\textbf{r}_2;\textbf{r}_1)^2
\end{eqnarray}
and $Dz\equiv dz \frac{e^{-z^2/2}}{\sqrt{2 \pi}}$; while $
v(\textbf{r})=G[\textbf{r};\Gamma]\equiv
\hat{G}[\hat{h}(\textbf{r});\Gamma]$ is the self-consistent
solution of
$v(\textbf{r})=F(\hat{h}(\textbf{r})+\Gamma(\textbf{r})v(\textbf{r}))$,
and, finally, $\hat{h}(\textbf{r})\equiv
h(\textbf{r})-\Gamma(\textbf{r})v(\textbf{r})$ is the part of the
local field at $\textbf{r}$ which does not directly depend on
$v(\textbf{r})$.

If spatially modulated overlaps can be roughly described by the very first
Fourier modes, then writing the above equations in Fourier space
would help. For $\alpha\neq 0$ we focus, for simplicity, on a
network which lies on a $1D$ ring; the analysis could be extended,
though, to arbitrary dimension.

We write the connectivity matrix, $m$ and $\rho^2$ in their
Fourier modes and find the following fixed-point equations (from
now on $c(x) \equiv cos(\frac{\pi x}{L})$, $s(x) \equiv
sin(\frac{\pi x}{L})$ and $\Delta$ is a dummy function label that can be $c$,
representing a cosine, or $s$, representing a sine):
\begin{eqnarray}
\wp(r_2,r_1)&=&\sum_{k} \tilde{\wp}_{k}c(k(r_2-r_1))\nonumber\\
\tilde{m}_{k}&=&\frac{\tilde{\wp}_{k}}{C}\int dr c(kr) I_2(r)\nonumber\\
\tilde{\rho^2}_{k}&=&\frac{\alpha T_0^2}{(1+\delta_{k0})LC}(Y^{k}_1+Y^{k}_2+Y^{k}_3)\nonumber\\
 \psi(r_2;r_1)&=&\sum_{kj \Delta}\psi^{\Delta}_{kj}\Delta(kr_2)\Delta(jr_1)
\end{eqnarray}
where we have:
\begin{eqnarray}
\psi^{\Delta}_{kl}&=&\tilde{\wp}_{k}\tilde{\wp}_{l}\sum_{n} \Phi^{\Delta,n}_{kl}\nonumber\\
\Phi^{\Delta,n+1}_{kl}&=&\sum_{i} \tilde{\wp}_{i}\Phi^{\Delta,n}_{ki}\Phi^{\Delta,1}_{il}\nonumber\\
\Phi^{\Delta,1}_{ij}&=&\int dr\Phi(r) \Delta(ir) \Delta(jr)
\label{psi_def}
\end{eqnarray}
and:
 \begin{eqnarray}
Y^{k}_1&=& (1+\delta_{k0})L  \tilde{\wp}_{k} \int dr c(k r)I_3(r)\nonumber\\
Y^{k}_2&=&2\sum_{\Delta}\int dr Q^{\Delta}_{k}(r)I_3(r)\nonumber\\
Y^{k}_3&=&\sum_{\Delta}\int dr W^{\Delta}_{k}(r)I_3(r)\nonumber\\
Q^{\Delta}_{k}(r)&=&\sum_{ijl}\tilde{\wp}_{i}\psi^{\Delta}_{jl}\Delta(ir)\Delta(lr) \Pi^{\Delta}_{ijk}\nonumber\\
W^{\Delta}_{k}(r)&=&\sum_{ii'jl}\psi^{\Delta}_{ii'}\psi^{\Delta}_{jl}\Delta(i'r)\Delta(lr)\Pi^{\Delta}_{ijk}\nonumber\\
\Pi^{\Delta}_{ijk}&=&\int dr\Delta(ir)\Delta(jr)c(kr).
\end{eqnarray}

Assume that the $b$ term in Eq.$\ref{locfield}$
depends only on the mean activity, and it keeps it constant and equal to $a$. The above equations for
$\tilde{m}_{r},\tilde{\rho}_{r},\tilde{\psi}^{c}_{r},...$ can
now be solved iteratively. The number of terms that one includes
in the sum in Eq.\ref{psi_def} (defining $\psi^{\Delta}_{rs}$),
together with the number of modes that one considers to
approximate the connectivity structure, determine the accuracy of
the calculation.

\begin{figure}[h]
\centerline{\hbox{\epsfig{figure=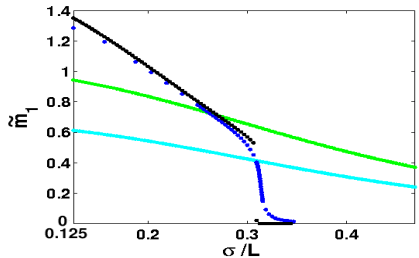,height=6cm,width=3.6cm,angle=270}}}
\caption{$\tilde{m}_1$ versus $\sigma$ for $g=0.5$, {\em TL} units
and $\alpha=0$ (black) and $\alpha=0.1$ (blue). The cyan curve is
for $0-1.5$ binary units and green for $0-2$ {\em B} units, both
at $\alpha=0$.} \label{fig1}
\end{figure}

Let us now concentrate on two specific input-output functions. For
the {\em TL} and the binary ($B$) transfer functions we have:
\begin{eqnarray}
G[r,\Gamma]_{TL}&=&\frac{g}{1-g\Gamma}(\hat{h}(r)-Th)\Theta(\hat{h}(r)-Th)\nonumber\\
G[r,\Gamma]_{B}&=&\Theta(\hat{h}(r)).
\end{eqnarray}
To proceed further we also assume a Gaussian connectivity:
$\wp(r_2-r_1)=(C/\sqrt{2\pi\sigma^2})\exp[-(r_2-r_1)^2/2\pi\sigma^2]$.
In Fig.\ref{fig1} we plot the amplitude of the first Fourier mode
$\tilde{m}_1$ -- an indication of the deviation from the uniform
solution -- as a function of $\sigma$ for {\em TL} units (black
$\alpha=0$, blue $\alpha=0.1$; in this figure and others always
$a=0.2$ and $C/N=0.05$) and for {\em B} units (cyan and green
curves, see below). With {\em TL} units, for small enough
$\sigma/L$ the solution is essentially a localized bump. One can
also see that decreasing quenched noise changes the transition to
non-uniform retrieval from a smooth to an abrupt one.

$0-1$ binary networks fail to exhibit non-uniform retrieval
(Fig.\ref{fig1}; \cite{Ana05} and \cite{Kor05}). The reason is
simply related to setting the sparsity, and it can be understood
intuitively as follows. There are two conditions to be satisfied
for the retrieval state to exist: $m_0=\frac{1}{N}\int dr
(\eta^{\mu}(r)/a-1) v(r)=O(1)$ and $x=\frac{1}{N}\int dr v(r)=a$.
The second means that, for spatially modulated retrieval states,
in some parts of the network units with activity 1 in the
corresponding stored pattern should have activity below 1, and in
other parts above 1. The latter requirement poses no problem to
the {\em TL} network, whose units can reach high levels of
activity. For a network with binary units, or with units that
saturate, the crucial issue is whether the {\em up} state, or the
saturation level, is sufficiently above 1 (the arbitrarily set
activity level of active units in the stored patterns; obviously
the argument can be generalized to non-binary stored patterns).
Thus binary units with activity levels, say, $0$ and $1.5$
(relative to the {\em up} state in the stored pattern) should be
able to show spatially modulated activity profiles, although,
rather than localized {\em bumps}, they appear as square-shaped
spatially restricted activity. This results in the cyan and green
curves for $\tilde{m}_1$ in Fig.\ref{fig1}.

To further assess the effect of the saturation level on the formation
of localized retrieval states we consider the following
input-output function:
\begin{equation}
F(x)=\varepsilon \tanh(gx/\varepsilon)\Theta(x)\nonumber.
\end{equation}
One should notice that $g$ is the slope at threshold and that for
a sufficiently high $\varepsilon$ this is effectively just a {\em
TL} function. For simplicity we focus on the $\alpha \to 0$ limit, as we
do not expect the quenched noise to make any qualitative change in
the behavior of the system, except for the smoothness of the
transition. Fig.2 shows how $\tilde{m}_1$ changes with $\sigma$
for fixed $g=0.5$ and different values of saturation, as measured
by $\varepsilon$. When the saturation is set at $\varepsilon=1$,
for the intuitive reason above the first Fourier mode does not
differ from zero. By increasing $\varepsilon$, however, one
approaches the {\em TL} regime. In Fig.3 we plot $\tilde{m}_1$
versus $\sigma$ for different values of $g$ and both {\em TL} and
saturating input-output functions. Notice the quasi-linear
behavior for values of $\sigma$ below the transition.
\begin{figure}[h]
\centerline{\hbox{\epsfig{figure=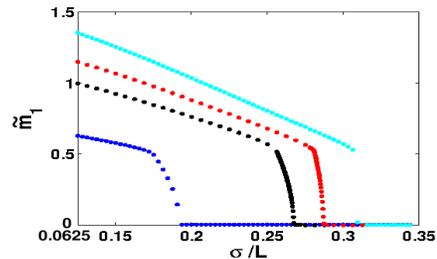,height=6cm,width=3.6cm,angle=270}}}
\caption{$\tilde{m}_1$ versus $\sigma$ for $g=0.5$ for different
values of the saturation level: $\varepsilon=2$ (blue),
$\varepsilon=3$ (black) and $\varepsilon=4$ (red). Cyan: {\em TL}
units ($\varepsilon \to\infty$).} \label{fig2}
\end{figure}

\begin{figure}[h]
\centerline{\hbox{\epsfig{figure=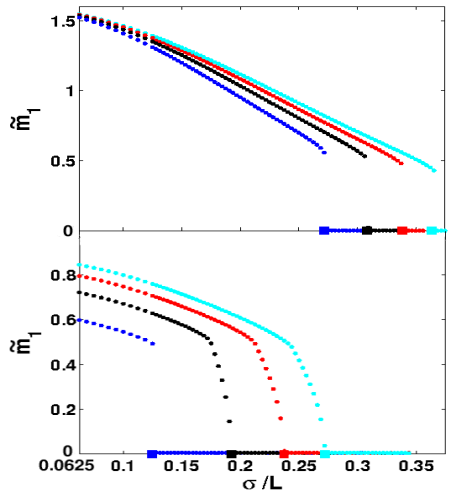,height=6cm,width=6.5cm,angle=270}}}
\caption{$\tilde{m}_1$ versus $\sigma$ for different values of the
gain: $g=0.45$ blue, $g=0.5$ black, $g=0.55$ red and $g=0.6$ cyan.
(upper panel) {\em TL} and (lower panel) saturating units with
$\varepsilon=2$. With saturating units decreasing $g$ (thus
linearizing the input-output function close to threshold) sharpens
the transition. The filled squares represent $\sigma_c$ as
predicted by Eq.\ref{sig_c}.} \label{fig3}
\end{figure}

When $\alpha=0$, analyzing the formation of non-uniform solutions
becomes simple even for a $d$-dimensional network. In this case,
the fixed-point equations read:
\begin{eqnarray}
\frac{\tilde{m}_{{i_1},\dots{i_d}}}{\tilde{\wp}_{{i_1},\dots{i_d}}}&=&\frac{1}{C}\int
d\textbf{r}
\prod^{d}_{n=1}c(i_nr_n)<(\frac{\eta^{\mu}(\textbf{r})}{a}-1)F(h(\textbf{r},\eta^{\mu}))>\nonumber\\
x&=&\frac{1}{N}\int d\textbf{r}
<F(h(\textbf{r},\eta^{\mu}))>\nonumber\\
h(\textbf{r},\eta^{\mu})&=&(\frac{\eta^{\mu}(\textbf{r})}{a}-1)m(\textbf{r})-Th.
\end{eqnarray}
One can see that
$m_{{i_1},{i_2},\dots{i_d}}=(1-a)\delta_{i_{1}0}\delta_{i_{2}0}\dots\delta_{i_{d}0}
$ and $Th=(1-a)(1/a-1)-F^{-1}(1)$ solve the above equations,
provided $F^{-1}(1)<(1-a)/a$ \cite{note1}. However, for the
following connectivity probability distribution:
\begin{equation}
\wp(\textbf{r},\textbf{r}')=\frac{c}{(2\pi\sigma^2)^\frac{d}{2}}\prod^{d}_{i=1}
exp(\frac{-(r_i-r'_i)^2}{2\pi\sigma^2})\nonumber
\end{equation}
this solution is stable only for $\sigma>\sigma_c$, where:
\begin{equation}
\sigma_c=\frac{L}{\pi}\sqrt{\frac{2}{d}\ln(a(1/a-1)^2F'(F^{-1}(1)))}
\label{sig_c}
\end{equation}
and $L$ is the half length of each dimension. For
$\sigma<\sigma_c$ the uniform solution becomes unstable in the
direction of the first Fourier mode $\tilde{m}_{1\dots 1}$. In a
network of  {\em TL} units the equation for $\tilde{m}_{1\dots 1}$
-- provided $\tilde{m}_{1\dots 1}<a/[g(1-a)]$ -- reads
$\tilde{m}_{1\dots 1}=g(\tilde{\wp}_{1\dots 1}/C)a(1/a-1)^2
\tilde{m}_{1\dots 1}$, which at $\sigma_c$ is satisfied for any
$\tilde{m}_{1\dots 1}$. This means that the system is marginally
stable at this point in the direction of $\tilde{m}_{1\dots 1}$,
resulting in a jump to $\tilde{m}_{1\dots 1}=a/[g(1-a)]$ for
$\sigma<\sigma_c$, as shown in the graphs. It is worth noting that
such trivial equation for $\tilde{m}_{1\dots 1}$ comes directly
from the linear nature of the {\em TL} function above threshold.
Adding quenched noise to a network of {\em TL} units or using any
other function, e.g. the 2 introduced above, would change this
trivial equation to a non-linear form, with the disappearance of
the jump, as evident in figures Fig.\ref{fig1}, \ref{fig2}, and
\ref{fig3}.

\begin{figure}[h]
\centerline{\hbox{\epsfig{figure=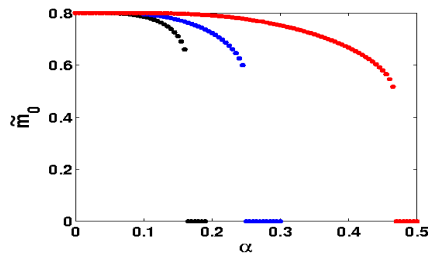,height=6cm,width=3.5cm,angle=270}}}
\caption{$\tilde{m}_0$ versus $\alpha$ for different values of
$\sigma$ in a {\em TL} network with $g=0.5$. Black for
$\sigma=500$, blue for $\sigma=800$ and red for structure-less
network. Note the corresponding values: $\tilde{m}_1=1.22$ for
$\sigma=500$, $\tilde{m}_1=0.8$ for $\sigma=800$ and
$\tilde{m}_1=0$ for the structure-less network, all when
$\alpha=0.05$.} \label{fig4}
\end{figure}

We further investigate the effect of localized retrieval on the
storage capacity of the network. In Fig.\ref{fig4} we plot
$\tilde{m}_0$ as a function of $\alpha$ for $\sigma=500$ and
$800$. Even though the storage capacity ($\alpha_c={\rm
Inf}_{\alpha}\{\alpha | \tilde{m}_0(\alpha)=0\}$) decreases, the
decrease is not too severe, even for very localized solutions.
Thus, the maximum number of retrievable patterns remains
proportional to $C$.

The results presented here show that in general, a network with
realistic single unit input-output transfer function becomes
capable of localized retrieval simply by manipulating single unit
saturation and linear gain. Increasing the gain, and/or the
saturation level, makes retrieval states more localized. These
parameters can be effectively controlled via inhibitory
mechanisms. The effect of the quenched noise is minor on the
qualitative behavior of the system, making the analytic formula
Eq.\ref{sig_c} a reasonable approximation for a wide range of
parameters. The fact that for a given value of $\sigma$, changing
the saturation level or the slope at threshold ($g$) can put the
network out of the spatially modulated retrieval regime may
explain the result of \cite{Ana05}.

Localized retrieval, while quantitatively decreasing local storage
capacity, may considerably increase the computational power of a
network with structured connectivity. This can be appreciated by
noting that in a large network, more than one memory pattern of
activity may be retrieved at the same time, each in a different
location, without much interference. A combination of locally
retrieved memories can be thought of as a global, composite memory
pattern. The number of such composite patterns would be
combinatorially large, thus hugely increasing the overall storage
capacity of the network.

Moreover, while sparsity measures e.g. in IT cortex tend to
yield high values (such as $a \simeq 0.7$ \cite{Rol98,Tre99}) -- seemingly
in contradiction with the notion that associative networks require
sparse coding in order to operate with a viable storage capacity --
our result suggests another perspective. The reported values of $a$
may have been measured, effectively, conditional to the recorded units
being active in a localized retrieval state, thus `overestimated' by
neglecting the large silent part of the network.

Each neuron in the neocortex receives of the order of $10^4$
synapses, and this number regulates a similar number of locally
retrievable patterns. The fact that the number of memories stored
in the neocortex seems much higher may stem from the combinatorial
character of global memory patterns, allowed by the localization
discussed here.

\end{document}